%
\font \twelverm=cmr10 at 12pt       
\font \twelvei=cmmi10 at 12pt       
\font \twelvesy=cmsy10 at 12pt      
\font \twelvebf=cmb10 at 12pt       

\def\ga{\gamma}

\def\lra{\longrightarrow}
\def\stars{\hfil\hbox to 1in{\leaders\hbox{ * }\hfil}\hfil}
\def\pattern{\hbox to 15pt{\hfil. \hfil}}
%




\def\gp{$\ga p \lra e^- e^+p$\/}
\def\ggg{$\ga \ga \lra e^- e^+$\/}
\smallskipamount=6pt  
\medskipamount=12pt   
\bigskipamount=24pt   

\font \sixteenrm=cmr10 at 16pt       
\font \sixteenit=cmti10 at 16pt      
\font \sixteeni=cmmi10 at 16pt       
\font \sixteensy=cmsy10 at 16pt      
\font \sixteensl=cmsl10 at 16pt      
\font \sixteenbf=cmb10 at 16pt       
%
%
\font \twelverm=cmr10 at 12pt       
\font \twelvei=cmmi10 at 12pt       
\font \twelvesy=cmsy10 at 12pt      
\font \twelvebf=cmb10 at 12pt       
%
\font \elevenrm=cmr10 at 11pt       
\font \elevenit=cmti10 at 11pt      
\font \eleveni=cmmi10 at 11pt       
\font \elevensy=cmsy10 at 11pt      
\font \elevensl=cmsl10 at 11pt      
\font \elevenbf=cmb10 at 11pt       
\font \eleventt=cmtt10 at 11pt      
\font \elevenbfit=cmbxti10 at 11pt  
\font \tenbfit=cmbxti10         
\font \ninerm=cmr9      
\font \ninei=cmmi9      
\font \ninesy=cmsy9     
\font \ninebf=cmbx9     
%
\font \eightrm=cmr8      
\font \eighti=cmmi8 \skewchar\eighti='177     
\font \eightsy=cmsy8 \skewchar\eightsy='60    
\font \eightbf=cmbx8     
%
\font \sixrm=cmr6        
\font \sixi=cmmi6   \skewchar\sixi='177        
\font \sixsy=cmsy6  \skewchar\sixsy='60       
\font \sixbf=cmbx6       
%

\def\ga{\gamma}

\def\lra{\longrightarrow}

%
%

%
                                                            
%

%
\def\sixteenpoint{\normalbaselineskip=16pt 
\def\rm{\fam0\sixteenrm}%
\def\it{\fam\itfam\sixteenit}%
\def\bf{\fam\bffam\sixteenbf}%
\def\sl{\fam\slfam\sixteensl}%
\def\tt{\fam\ttfam\sixteentt}%
\textfont0=\sixteenrm \scriptfont0=\twelverm
                                           \scriptscriptfont0=\elevenrm
\textfont1=\sixteeni \scriptfont1=\twelvei
                                           \scriptscriptfont1=\eleveni
\textfont2=\sixteensy \scriptfont2=\twelvesy
                                           \scriptscriptfont2=\elevensy
\textfont\itfam=\sixteenit
\textfont\slfam=\sixteensl
\textfont\bffam=\sixteenbf
\scriptfont\bffam=\twelvebf      \scriptscriptfont\bffam=\elevenbf
\def\doublespace{\baselineskip=32pt}
\def\singlespace{\baselineskip=16pt}
\def\openspace{\baselineskip=24pt}
\normalbaselines\rm}
%
%

%

%
\def\elevenpoint{\normalbaselineskip=12pt 
\abovedisplayskip 15pt plus 3pt minus 9pt 
\belowdisplayskip 15pt plus 3pt minus 9pt 
\abovedisplayshortskip 0pt plus 3pt 
\belowdisplayshortskip 9pt plus 3pt minus 4pt 
\def\secondsubheading{\elevenbfit} 
\def\rm{\fam0\elevenrm}%
\def\it{\fam\itfam\elevenit}%
\def\bf{\fam\bffam\elevenbf}%
\def\sl{\fam\slfam\elevensl}%
\def\tt{\fam\ttfam\eleventt}%
\textfont0=\elevenrm \scriptfont0=\ninerm  \scriptscriptfont0=\sevenrm
\textfont1=\eleveni \scriptfont1=\ninei  \scriptscriptfont1=\seveni
\textfont2=\elevensy \scriptfont2=\ninesy  \scriptscriptfont2=\sevensy
\textfont\itfam=\elevenit
\textfont\slfam=\elevensl
\textfont\bffam=\elevenbf
\scriptfont\bffam=\ninebf
\scriptscriptfont\bffam=\sevenbf
\def\doublespace{\baselineskip=22pt}
\def\singlespace{\baselineskip=12pt}
\normalbaselines\rm}
%
\def\tenpoint{\normalbaselineskip=12pt 
\abovedisplayskip 12pt plus 3pt minus 9pt 
\belowdisplayskip 12pt plus 3pt minus 9pt 
\abovedisplayshortskip 0pt plus 3pt 
\belowdisplayshortskip 7pt plus 3pt minus 4pt 
\def\secondsubheading{\tenbfit} 
\def\rm{\fam0\tenrm}%
\def\it{\fam\itfam\tenit}%
\def\bf{\fam\bffam\tenbf}%
\def\sl{\fam\slfam\tensl}%
\def\tt{\fam\ttfam\tentt}%
\textfont0=\tenrm \scriptfont0=\eightrm \scriptscriptfont0=\sixrm
\textfont1=\teni \scriptfont1=\eighti  \scriptscriptfont1=\sixi
\textfont2=\tensy \scriptfont2=\eightsy  \scriptscriptfont2=\sixsy
\textfont\itfam=\tenit
\textfont\slfam=\tensl
\textfont\bffam=\tenbf
\scriptfont\bffam=\eightbf
\scriptscriptfont\bffam=\sixbf
\def\doublespace{\baselineskip=20pt}
\def\singlespace{\baselineskip=12pt}
\normalbaselines\rm}
%

%

\def\gp{$\ga p \lra e^- e^+p$\/}
\def\ggg{$\ga \ga \lra e^- e^+$\/}

%

\voffset=.6in
\elevenpoint
\singlespace

\centerline{\sixteenpoint\sixteenbf 
A Word from a Black Female Relativistic Astrophysicist:}
\centerline{\sixteenpoint\sixteenbf Setting the Record Straight
on Black Holes}
\vglue .25in
\centerline{Reva Kay Williams}
\centerline{\it Department of Astronomy, University of Florida, 
Gainesville,
FL 32611}
\vglue .2in

\centerline{\bf ABSTRACT}
\medskip

\noindent
This {\elevenit Letter} is written to clear up a situation, 
and hopefully we will
learn something from it: scientifically and morally.  Herein is
presented a true ``historical'' scenario of events, that led to my 
being the first person  (Williams 1991) to successfully work out the Penrose
mechanism in  four-dimensions (three-space momenta and energy).
  Before working out a solution to the Penrose mechanism: to 
extract energy from a rotating black hole, the Penrose mechanism 
(since first proposed by Roger Penrose in 1969) had been attempted by 
scientists over the world for nearly two decades,  with little 
success, although making some progress. In the Penrose analysis of 
Williams (1991, 1995) details of the 
behavior of efficient Penrose relativistic scattering processes in the 
ergosphere are described. 
The reason a solution eluded other scientists before me is that there was 
very little known about orbits inside the ergosphere, where space and time 
are no longer separable, as measured by an observer at infinity (i.e., far 
away from the Kerr black hole).  In this {\elevenit Letter}, I describe how 
analytic derivations of the conserved energy and azimuthal angular momentum
of particle orbits not confined to the equatorial plane
allowed me to succeed where others, whose ``shoulders'' I
stood on, had failed. 
Also,  I will mention some well known scientists in the 
astrophysics 
community by name, as I discuss their
involvement, and outline the facts behind an author feeling the need to
set the record straight.  

\vfill\eject

\centerline{\bf I. INTRODUCTION}
\medskip

The problem is the following. There were, at least  
after May 1997 up to  January 1999,
two 
popular trains of thought associated with energy extraction and the
production of jets in black holes:
one is that the jets are inherent properties of geodesic
trajectories in the Kerr
metric of a rotating black hole, and thus,
can be described by Einstein's General Theory of Relativity
(Williams 1995; de Felice \& Carlotto 1997);
and the other
is that the accretion disk and its magnetic field through
magnetohydrodynamics (MHD) are producing the jets, with the field
being anchored to the rotating back hole or the accretion disk,
like models of Blandford \& Znajek (1977) and Blandford \&
Payne (1982).
My suggestion (Williams 2002) is that it could be a combination of the two,  
with
gravity  controlling the flow near the event horizon,
and MHD controlling the flow at distances farther away.
The observations of the jet of M87 suggest this may be
the case (Junor, Biretta, \& Livio 1999; Perlman et al.~2001). 

But, whatever the case maybe, the supporters of the
Blandford-Znajek (BZ) model of today, Blandford, in particular, and his 
collaborators, continue to promote this model,  even 
with the aged-old problem of converting from electromagnetic 
energy to relativistic particle energy, 
 not to mention the general relativistic  problems (see Williams
1999b and references therein).  
Now, I find nothing wrong with their continued support of  BZ-type
models. It is only
when these supporters publish researched work from my paper 
(Williams 1995) in 
which I describe in details gravitational physics near the event horizon;
yet, they  do not appropriately reference my manuscript: this is where 
the  problem lies.  It seems 
that they are
suppressing my successful,
competing model---that can extract energy from a black hole---by
masking my model, while using black hole 
physics 
originated from my manuscript, attempting  
to make BZ-type  models work.  Such behavior, of not references my 
work,  has resulted in
other
authors taking credit for black hole physics devised by me.

The latest attempt is Komissarov's recent paper  (astro-ph/0402403, 
accepted for
publication in MNRAS).  
This author claims that the driving force of
the BZ model is the ergosphere, just like in the  Penrose 
mechanism, if one assumes 
that the magnetic field lines are inertial frame dragged, 
generating an electric
field that accelerates (i.e., ``heats'') electrons that subsequently 
inverse Compton scatter soft background 
photon from the accretion disk.  {\elevenit Here is where I must 
speak out to set the
record straight!}  
I do not know exactly where Komissarov, who acknowledges ``Roger Blandford
for his useful comments on this subject,''  is going with this 
investigation. Since Komissarov has already established that the magnetic 
field need not anchor to the event horizon, but only be inside the 
ergosphere, which mildly 
suggests that the ``classical'' BZ model is not important, his direction 
is likely obvious and inevitable.  
This author goes 
further,  implicating, as stated above, 
that the generated  electric field heats the electrons 
that  inverse Compton scatter  soft background 
photon in the ergosphere.  Now, based on past experience, it seems
that Komissarov is being led, advised, or directed by Blandford, 
into eventually using my 
calculations (Williams 1995), i.e.,  first my 
Penrose Compton scattering in the ergosphere, then perhaps the rest:
Penrose pair production (\ggg), 
Penrose pair production (\gp); and, of course, with no intentions of 
referencing me
as the originator, as it is with some other 
authors associated with Blandford.  

Well, at this point, I must stop Komissarov and make him quite aware, 
as well as the
scientific community as a whole, i.e.,  any that may not know, 
 a complete four-dimensional spacetime analysis of
Penrose scattering processes (inverse Compton scattering and pair
production as mentioned above) has already being
thoroughly investigated by Williams (1995), inside the
ergosphere of a supermassive Kerr (rotating)
black hole.  In this analysis (1995; see also Williams 
1999a, 1999b, 2001, 2002),
general relativistic model calculations,
surprisingly, reveal
that the observed high energies and luminosities of quasars and other
active galactic nuclei, the collimated jets about the polar
axis, and the asymmetrical jets (which can be enhanced by
relativistic Doppler beaming effects) {\it all} are inherent properties
of rotating black holes. From this analysis,  it is shown
that the Penrose scattered escaping relativistic particles
exhibit tightly wound coil-like cone distributions (highly
collimated vortical jet distributions) about the polar
axis, with helical polar angles of escape varying from $0.5^o$
to $30^o$ for the highest energy particles.  It is  also shown that
the gravitomagnetic (GM) field, which causes the dragging of inertial
frames, exerts a force acting on the momentum vectors of the incident
and scattered particles, causing the particle emission to be asymmetrical
above and below the equatorial plane, thus appearing to break the
equatorial reflection symmetry of the Kerr metric. When the accretion disk
is assumed to be a two-temperature bistable thin disk/ion corona
(or torus $\equiv$ advection-dominated accretion flow),
energies as high as 54 GeV can be attained by these
Penrose processes alone;
and when relativistic beaming is included,
energies in the TeV range can be
achieved, agreeing with observations of some BL Lac objects
(Williams 1999b).  

These calculations of the Penrose mechanism (Williams  1995) have
already revealed in details how energy is extracted from
a black hole, suggesting a complete theory, (1) without the
need of a large-scale  accretion-disk magnetic field,
which, according to general relativity, will not effectively exist,
 upon nearing the event horizon; 
and (2) without the associated problems arising when attempting to use 
the disk magnetic
field
 in direct extraction of energy from a black hole.

So, what use is there for the BZ-type model proposed by 
Komissarov?  The Kerr rotating black hole gravitationally blueshifts 
particles to higher energies, so there is no need for this
to be done by the disk magnetic field, assuming it could 
effectively exist in 
the ergosphere, 
as proposed by  Komissarov, but contrary to recent findings 
that the flux of such a field will be expelled or redshifted away
(Bi\v{c}\'{a}k 2000;
Bi\v{c}\'{a}k \& Ledvinka 2000), relative to the observer at infinity
as well as locally.  

Moreover, it does not surprise me that an author, that has 
collaborated with Blandford, on some level, would  
state that 
the magnetic field is dragged around by spacetime in the 
ergospheric region.
 Notice the
similarity
to my proposing (Williams 2001, 1999a), for the first time ever, 
 that the field lines of the GM field
(i.e., the gravitational analog or resemblance
of a magnetic field)
inside the ergosphere are inertial frame dragged,  
causing symmetrical and
asymmetrical jet forces in the polar direction, which is 
based on detailed analytic calculations and computer 
simulated results.  
Note, I recently 
(April 2004) found in the literature
that Bi\v{c}\'{a}k \& Jani\v{s} (1985) were the first to propose  
magnetic field lines are frame dragged into rotation by the geometry,
inducing an electric field; why did Komissarov fail to reference 
this paper? 

This in general has 
happened a number of times, concerning my work.  It appears
that Blandford, down through the years, since our first contact
in 1992 in which he became familiar with my work
(as will be described in this {\elevenit Letter}), has constantly 
advised his collaborators and graduate students to do black hole
physics worked out and/or proposed  by me, yet not appropriately 
referencing the origin.  

Peer review
processes, associated with manuscript
submissions to {\it The Astrophysical Journal} (ApJ) and 
{\elevenit Physical Review D}  (PRD),  and  
research proposals submitted to
granting 
programs of NASA and the National Science Foundation (NSF), 
if dishonesty is involved, as it rarely
is, provide the perfect means for an author 
  to gain access to another author's unpublished works.  Those of
you who have served as referees or grant proposal panelists
know that it is all about the integrity of the individual, that keeps 
one from
wrongdoing.

Note, as an  originator 
of much of the ``modern''
black hole
physics of today, based on my 41 page paper
published in  PRD (Williams 1995), scientists at NASA's 
Astrophysics Theory Program,
NSF, and ApJ,
 have spoken highly of my work:
 She has contributed
significant to the field (NASA 1997).  She has impressive experience in
analyzing
rotating black holes, and has developed important insight into the
details of processes needed to liberate energy in the form of jets
(NSF 1999).  She is an expert on this kind of physics and her
three-dimensional analysis of the energy extraction process has
contributed to our understanding of the Penrose mechanism (NSF 1999).
Her work on the Penrose mechanism, and how one might use it in a 
quantitative manner to extract energy from within the ergosphere of
a spinning black hole using the solutions that she developed in the 
early 1990's, is first-rate (ApJ 2004).
            
So, now you are probably wondering, if Williams' model does
all that she claims, then why have we not heard more
of her work?  In the paragraphs below, I will outline the
chain of events, explaining how up until 1999, I was beginning to get 
international recognition from the astrophysics community for the 
black hole
physics introduced by me in Williams (1995), and  then it was 
abruptly taken away from me by a group of scientists
(collaborators, graduate students, friends, etc.~of Blandford).  
These scientists  began 
applying my devised  black hole physics to  BZ-type models,
but, as mentioned above,   were not
appropriately referencing my
paper.  This resulted  in these scientists being given credit for
my work.
 I am appealing to the scientific community to help me set the 
record straight. 

I always believed that I would be respected as a Black person for my
knowledge.  This was one of the reasons I was determined to get my PhD, and 
with me being the first Black female astrophysicist in our nation, 
it was not easy, yet I pressed through.
Now it seems that my knowledge is being unethically used against me,
by some authors, for selfish and economical gains.  That is, by
not 
appropriately referencing me as the originator of researched
black hole physics presented 
in such a classic, important paper (Williams 1995) has 
resulted in certain authors taking 
credit for my research, and thus, causing my career to suffer. 
  Is it because I am Black,  or because I am
a woman, or both?
                  
In the following sections, the events leading to the problem stated above 
and people involved will be
presented, alone with documentation when needed for
credence and/or  clarity.  
\vglue .5in

\centerline{\bf II. GRADUATE SCHOOL}
\medskip

While studying physics as an undergraduate, I realized early on that
my two favorite forces are gravity and electromagnetism.  So, while in
graduate school, every chance I
got, I did a paper, gave a talk, or worked with some faculty person
on a research project having to do with one of these
forces, whether it was studying the magnetic field structure of disk 
galaxies, magnetic fields in jets, magnetic fields of neutron stars 
and white dwarfs, or just working problems in 
electromagnetism for a summer
research project---I was always eager to learn more so that I could
apply the knowledge to solving problems of the universe related 
to these forces.
Since, however, done of the faculty members at Indiana University, 
Bloomington,
were  doing relativistic astrophysics research, I 
taught myself General and Special Relativity at an advanced level,
from many books, including Weinberg (1972) and scientific
journal papers. 

Because I was not working on any of my advisor's grants, I had to work and
save money to attend conferences on my own.
I attended the 1988 Texas Symposium on Relativistic Astrophysics, 
hoping to meet 
Penrose and some of the other conferential scientists, including those 
whose work I had
become familiar with through their journal papers and correspondence,
such as Williams Fowler,  Blandford, Martin Rees, Stephen Hawking.
My mission was to find out  why, it seemed, no one was working on
the Penrose mechanism. Yet, I would still see it mentioned in text books,
as if people still believed in its potential.
But I could not find anywhere in the literature of it being worked out.

Below, I give a brief chronological account of my research projects 
in graduate
school, as a Research Associate (1984-1991), before getting my PhD:
\bigskip
\itemitem{}{\bf 1984-1985:}
This marks the beginning of my dissertation research.  My desire was
to created a model to explain the core energy source of quasars and other
active galactic nuclei (AGNs). Next, I desired to use this energy source to
generate the jets associated with radio strong AGNs.
To begin with, I duplicated William Fowler's supermassive star
($\sim 10^8 M_\odot$) model computer simulation of a $n=3$ polytrope as a
possible energy source of quasars.  A problem encountered in using these
objects is that the life expectancy of supermassive stars shining on
nuclear energy is only $\sim 10^6$~years. Quasars are thought to have ages
up to $ 10^9$~years: supermassive black hole models with accretion
disks can supply the necessary energy for at least $10^9$~years; however,
supermassive stars could very well be progenitors of such black holes
(Fowler 1985; by correspondence).
\bigskip                                                              
\itemitem{}{\bf 1986-1987}:
I  computed a model simulation of a thin disk/ion corona accretion,
surrounding a Kerr black hole,
with accretion rate $\dot M < \dot M_{\rm Eddington}$.
This model combined the thin accretion disk of
Novikov and Thorne (1972) with
 the two temperature ion corona (or thick disk) of
Eardley and Lightman (1975); 
Eilek (1980); Eilek and Kafatos (1983).
This computer generated model readily gives parameters of the
disk, such as number density, pressure, temperature, height, etc.,
at a given radius, before and after instability sets in, i.e., the
secular instability that
makes the thin disk swell into a thin disk/ion corona bistable geometrical
structure.  The problem one encounters here is that the accretion disk
can only give rise to particle energies up to $\sim 100$~MeV (Eilek 1980).
Yet quasars are observed to radiate energies up to at least $\sim$~GeV.
So the only place I had to turn to, hoping to get the high energy needed
to duplicate the observed spectra of quasars, was the black hole.
\bigskip                                                                
\itemitem{}{\bf 1987-1990}:
I computed a Monte Carlo simulation of Compton scattering in the ergosphere
of a supermassive Kerr (rotating) black hole, using the Penrose mechanism
to extract rotational energy.  A similar calculation was previously done by
Piran and Shaham in 1976; however, they applied the equatorial
escape conditions, i.e., for photons confined to
escape along the equatorial plane, thus, ignoring the
escape conditions for scattered photons above and below the plane, assuming
that the equatorial escape conditions were not much different from 
the nonequatorial escape conditions.  Yet, I found that the
escape conditions were very different,
thus, making their assumption invalid.
In my model calculation, I apply the
nonequatorial escape conditions
to the scattered photons
(i.e., escape conditions that take into
account motion above and below the equatorial plane).
However to do this, I had to derive analytical solutions for
the conserved energy and angular momentum
of nonequatorially confined orbits of massless and material 
particles; such solutions
had never been derived before.
These derived solutions allowed me to apply the nonequatorial escape 
conditions, from which I obtained higher energies and more photons 
escaping than
obtained by Piran and Shaham.
The results of my model calculation show that the Penrose Compton scattering
can generate  hard X-ray/soft $\ga$-ray spectra in agreement with
observations of AGNs.  But more so, this black hole model calculation
can incorporate
other scattering processes that could generate even higher energies.
The results
of the above investigation was presented in my thesis [Williams, R.~K.,
$Master's~Thesis$, Indiana University, Bloomington (1990)].
\bigskip
\itemitem{}{\bf 1991}:
For my dissertation (December 1991),
I presented Monte Carlo computer simulations
of Penrose Compton scattering (PCS) and Penrose pair production
production (PPP) processes, \gp\space and \ggg,
in the ergosphere of
a supermassive ($10^8 M_\odot$) rotating black hole.
This mechanism, as
applied in these calculations, can extract
hard X-ray/$\ga$-ray photons from
the inverse Compton scatterings of initially low energy
UV/soft X-ray photons by target orbiting electrons in the ergosphere.
Such low energy infalling
photons are consistent with photons emitted by that of a
thin disk/ion corona accretion. These model calculations also allow
relativistic electron-positron ($e^-e^+$)  pairs
to escape
with energies as high as $\sim 4$~GeV: these pairs are
produced by
infalling low energy photons interacting with target photons in bound
orbits inside the ergosphere, at the {\elevenit photon orbit}.
This process may be the origin of the relativistic electrons
inferred from observations to emerge from the cores of AGNs.
Also discovered in these model calculations were that the Penrose
scattering processes naturally produces jets of relativistic particles
aimed toward the polar axes, and in most cases the jets are
somewhat one-sided,
agreeing with observations of AGNs.
Overall, these  Penrose processes can apply to any mass black hole,
more or less, depending on the characteristics of the accretion disk,
and suggest a complete theory for the extraction
of energy from a black hole.
[Note that, the  above investigation was subsequently published in
{\it Physical~Review~D}, 51, 5387 (1995).]
\bigskip

\centerline{\bf III. POSTDOCTORAL ERA}
\medskip

As I applied for postdoctoral fellowships, 
I sent my Proposed Plan of Research (1992)
to various institutions.  On one occasion, I 
sent, along with the Proposed Plan of Research, a copy of my dissertation. 
 This was sent to Kip Thorne at California Institute of Technology
(Caltech).  This occurred after
applying for a postdoctoral position at the Space Telescope Science 
Institute, and being told my Meg Urry in November 1992 
that she would be glad to give me a 
position, but it would be observing.  She said that I needed to be in
a position where 
I could do theory.  She then suggested that I contact Martin
Rees or Kip Thorne.  I was quite honored, and replied that I never
thought about contacting these scientists of such great status.  She
then replied, ``but you are good.''  Her remarks were based on comments
told to her by Jean Eilek, her friend, who was also on my dissertation
committee.  

Since, I had already made plans to attend the 1992 Texas Symposium on
 Relativistic
Astrophysics, held in Berkeley, CA, to present my research
and to meet with Penrose, giving him a copy of my dissertation, telling
him that ``it worked,'' and  since Kip Thorne
was on the conference committee,  figuring that I could talk with him at
the meeting, I sent him a copy of my dissertation and my 
Proposed Plan of Research, inquiring about doing 
a Ford Foundation
Postdoctoral Fellow at Caltech.

In the Proposed Plan of Research, I proposed to 
investigate Penrose processes in the electromagnetic field of the Kerr-Newman
(rotating and charged) black hole (see also Williams 1995, p.~5423), 
repeating
the calculations of the Penrose scattering processes,
that were initially done  in the field of a rotating Kerr black hole.   
Since NASA's Gamma Ray Observatory (GRO) 
had recently confirmed that 
quasars (at least 3C~273 and 3C~279) radiated up to $\sim 10$~GeV, and my 
model at the time, based on limited knowledge of the accretion disk,
extended only up to $\sim 4$~GeV, but could extend higher in an
appropriate magnetic field, plus I needed a way to generate the 
electromagnetic jets: to accelerate and collimate the Penrose escaping
particles, led me to propose using the Kerr-Newman black hole.  
That is, I had found a way to extract high energy particles;
now I was ready to create an electrodynamic model to generate the
jets  (proceeding on with my initial plan made in graduate 
school).  I had proposed to let the 
magnetic field of the Kerr-Newman black hole
interact with the Penrose escaping electron-positron ($e^-e^+$) 
pairs.
From a purely electrodynamic point of view, I proposed to 
have the Keplerian 
conducting accretion disk move through the dipolar-like magnetic 
field of the black hole, inducing an electric field, similar to that of 
Lovelace (1976) and Blandford (1976), but in my case the high energy
particles would already be available, and not have to be created in some
complicated improbable way in the field, as proposed in the above 1976 
models.  Also, since my disk would be treaded by the magnetic field of the 
black hole ($B\sim 10^{10}$~gauss), as opposed to the field of the 
accretion disk (up to $B\sim 10^{4}$~gauss), a more powerful jet 
would be expected.  Finally, I proposed to
compare my results with some of the "opposing" models
that proposed using the electromagnetic field of the accretion disk
(Blandford 1976, Blandford \& Znajek 1977;
Blandford \& Payne 1982; Lovelace 1976, Scott \& Lovelace 
1982; Burns \& Lovelace 1982; Punsly 1991)---as opposed to
using the electromagnetic field of the Kerr-Newman black hole---to generate
the astrophysical jets of AGNs.                 

Note, in the Proposed Plan of Research,
I added, stating, ``From this project, I will receive the necessary background
knowledge needed in order to begin a successful research career,
studying general relativistic astrophysics and electromagnetic~fields
of massive extragalactic bodies. I will also apply this
knowledge to phenomena in cosmology---especially to phenomena
pertaining to the early universe.''         

I arrived in California at the 1992 Texas Symposium on Relativistic
Astrophysics  meeting on a Sunday, and would present a poster 
paper on the following Wednesday.  I found Penrose right away.  I had planned 
to meet Thorne and Blandford also.  Now,  Thorne was okay, i.e., as 
far as his research goes.  But Blandford was the scientist of whose work I
had investigated in details, in my quest to find the ``perfect'' quasar model.

Deviating briefly,
Blandford was one of the astrophysicists, I had concluded, whose
model came close  
to getting it right, he and Lovelace.  While in graduate school, I studied
both their models, for the generation of jets by electrodynamics: The 
Lovelace type models
(Lovelace 1976, Scott \& Lovelace 1982; Burns \& Lovelace 1982)
assume $e^-e^+$ cascades are accelerated by the electric 
field, generated from a magnetized rotating disk.
The Blandford type models (Blandford 1976, Blandford \& Znajek 1977; 
Blandford \& Payne 1982) assume the disk 
magnetic field anchors to the 
event horizon, i.e., ``surface'' of the black hole, extracting
electromagnetic energy due to magnetic braking, or anchors to
an  accretion, such that centrifugal forces 
 launch or lift accretion disk particles to relativistic 
speeds.  
But the main problem with both these jet models was converting
from Poynting flux of electromagnetic energy-momentum to
relativistic particle flux energy-momentum.  In both cases, the magnetic 
field had to be unphysically large or the particle density
had to be unphysically small.  
 In essence, Blandford and Lovelace needed a way to get the
high energy particles, that could subsequently interact with 
their proposed 
electromagnetic
field, to generate the jets.  So, standing on the ``shoulders''
of Blandford and Lovelace,
I figured I could use their way to accelerate and collimate the 
particles---of course, appropriately referencing their work;
but first, I had to get the relativistic particles
needed. 
My main focus became, at that point, to find a way to
get the $e^-e^+$ pairs.  

After searching the literature, 
I was led to the Penrose mechanism
by some authors (Piran, Shaham, \& Katz 1975; 
Leiter \& Kafatos 1978; 
Kafatos \& Leiter 1979; Kafatos 1980; Piran \& Shaham 1977a,
1977b; Eilek \& Kafatos 1983) who came close to working it out.   
But different from them,  however,
standing on their ``shoulders,'' I was able to
 extract high energy photons, and most
importantly, to extract relativistic $e^-e^+$ pairs in the
form of jets, thus, working out the Penrose mechanism.

Note,
my solving the orbital equations for the conserved energy $E$
and azimuthal angular momentum $L$ (Williams 1991, 1995), of
nonequatorially confined particle trajectories, above and below
the equatorial plane, allowed me to do the PPP (\ggg)
 at the photon
orbit.  This is because the bound photon can only exist in
nonequatorially
confined orbits in the Kerr metric.

So, as mentioned earlier, 
for my postdoctoral fellowship, I proposed to use the magnetic
field of the Kerr-Newman black hole, comparing it to that of the 
accretion disk, to further accelerate and  collimate the Penrose
produced escaping particle, expectedly, out 
to the observed distances.  It did not matter to me which would be
more important, the field of the black hole or that of the accretion
disk:  I just wanted to get to the scientific truth.
When I wrote to  Thorne, to inquire about coming to Caltech
as a postdoctoral fellow, I thought he too wanted to know the
truth, for I knew that I had part of it [i.e., the relativistic
$e^-e^+$ pairs from PPP (\ggg) at the photon orbit].  

Below I present exerts from an email letter sent to me from 
Thorne in response to my inquiry concerning the Ford 
Postdoctoral Fellowship. 
But first I must return to mentioning the occurrences at the 1992 Texas 
Symposium surrounding my meeting Thorne.  After working out the Penrose 
mechanism in 1991, showing that the energy observed astronomically can be 
 extracted from rotating black holes, and thus, placing them at
the cores of quasars, in 1992 NASA's GRO 
began confirming my model.  But, when Carl Fichtel of
NASA gave a talk that Tuesday on the new high energy GRO 
results, 
towards the end
of his talk,  he made the statement, along with showing a schematic
drawing, that recently Kip Thorne and his group (with my name not mentioned)
 suggested to use the magnetic field of the black hole!
This was my proposal that I had sent to Thorne a few weeks earlier.

I was literally crushed.  On my way out of the auditorium, almost
in tears, I met Janna Levin, a then graduate student at MIT, and 
expressed to 
her what had taken place.
She try consoling me because the evening before, at dinner, we,
and some others, including her research advisor, had discussed
such unethical behavior.
I then looked for Thorne, but was told that he would not be
at the meeting until Friday.

During my poster presentation the following day, I told everyone 
that I could what happened (i.e., the fact that my model had been 
mentioned by Fichtel), in my desire, now, to find just what
type of person was Thorne, for I had never met him personally.
The feedback was not very encouraging.  I even told it to Meg Urry, 
who also attended the meeting.  She thought it was a good thing,
and that I could work with Thorne on the project. 
I also heard that Thorne was at odds with the astronomy community
because of the large funds he was taking from the 
granting agencies to complete the LIGO project.  I figured, okay, if he 
just wanted to ``borrow'' my idea, just so that the community
would  think that Kip Thorne is still in the business of new ideas, then fine.
However as you will see from the letter exerts below,
Thorne's intentions were quite contrary to what Urry and I thought.

Finally, when Thorne arrived at the meeting on Friday, someone
came over and pointed him out to me.   I went up to him and introduced 
myself.  He responded that he would talk with me after doing lunch 
with Stephen Hawkin.   
 
After his lunch, while I was talking with Janna,  Joseph Silk, 
and another
scientist, Thorne came up to me stating he was ready to talk.
He then took me around the corner away from the crowd.  I immediately
asked him what did he think of my proposing to use
the magnetic field of the Kerr-Newman black hole.  His response was that 
he had not gotten to that, but, he was just so interested in my 
cover letter.
I said, ``But you had to, because Carl Fichtel mentioned that Kip
Thorne and his group suggested to use the magnetic field of the 
black hole.''  He responded that he was not there, and therefore, 
he did not know 
what Fichtel said.  When I saw Thorne appearing to play
the ``I don't know role,'' I backed off.   I surely did not want 
a man of his status to come down on me (a person just beginning
my career).  So, I changed the subject,  asking what I would need 
in regards to my application package for a postdoctoral position at
Caltech.  We talked more about the application, and he gave me his email
address.

When I got back home to Indiana University, I wrote Thorne a
letter, telling him the story of my long efforts,
i.e., my long hours spent and years of research to find a
model that explained quasars.   I expressed to him that I was almost 
there: I had the particles needed to interact with a surrounding 
electromagnetic field, that could possibly  generate the jets out to 
Kpc distances.
I wanted to convey to him of my being a real person, and not just 
something you
can, or would even want to, take or steal from.

I continued to complete my Ford Foundation Postdoctoral Fellowship 
application, with Caltech as the  host university.  After completing my 
application
all but the host letter from Thorne, and the deadline having passed,
although the office was willing to make an exception, I contact Thorne
to remind him.  Below are exerts from the email of February 25, 1993
Thorne sent to me:
\bigskip

{\tenpoint
\narrower\narrower
\noindent Dear Reva [Kay],

\smallskip
I am replying$\ldots$to your application to come to 
Caltech on a temporary research position, with your own salary support.
The bottom line is positive, but with a caveat.

Roger Blandford, Sterl Phinney, and I have all read the letters of
recommendation that you arranged to be sent, and your research 
proposal and portions of your thesis.  On these basis, we would 
be happy to have you as a member of our research group next year,
and would look forward to interacting with you.  however, none of
us is enthusiastic about your proposed continued research on the 
Penrose process as an energy source in AGNs.  We (particularly Blandford 
and Phinney, who are much better experts on this than I) 
believe this process is NOT very promising.  We also, quite generally,
believe that postdoctoral researchers should branch out in new
directions, different from their thesis work; this is important
if they are to be strong scientists.

Accordingly, we would be happy to accept you into our group if you
will agree that you will spend the majority of your research and 
writing time on new research directions, different from the 
Penrose process, directions to be selected  in consultation with
Blandford, Phinney, and me.  (I would encourage you, however, to 
finish writing up your past Penrose process research in parallel with
launching  one or more new research directions.)

Your new research directions could concentrate on relativity$\ldots$.
I would be especially interested in seeing you get involved with issues 
in the generation of gravitational waves by astrophysical systems
(e.g. coalescing binaries) in connection with the LIGO Project$\ldots$.

Please let me know your reaction to this.  If it seems acceptable to
you, then I will write to the$\ldots$Fellowship program$\ldots$.

\smallskip
\noindent Best wishes,

\noindent ~~~~~~~~~~~~Kip
}
\bigskip

When I showed this letter to one of my dissertation
advisors, he did not understand 
why they would advise me not to finish 
my initial proposed project (which consisted of two parts) 
as a postdoctoral fellow, in addition to finish 
writing up ``Part~1,''  especially since I had had so much 
success 
with Part~1. (Part~1 was
to extract the relativistic particles, particularly
the $e^-e^+$ pairs; and ``Part~2'' was to use
the surrounding magnetic field, interacting with these particles,
to generate the jets.
My advisors thought that the completion of Part~1 was 
enough to get my PhD, calling it ``a very fine piece of work
and I am sure she will continue to expand fruitfully upon it'';
it is  ``beyond the ability of many of her 
contemporaries''; ``a  first-class piece of work.'')
Other than that, one might think that the letter above
is not so bad.  However, considering what had taken place 
at the Texas Symposium, a few weeks earlier, i.e., how
Fichtel announced that Kip Thorne and his group suggested 
using the magnetic field of the black hole, one can not help
but conclude that something is amiss.  If Thorne and his
group at Caltech 
were proposing to use the magnetic field of the black
hole, independent of my proposal, then why would they
not want me to work on the same model with them, 
since I had devised a theoretical and numerical model 
to extract the relativistic particles that would be needed?

The reason why, it seems, is that Thorne and his group, including 
Blandford and Phinney, wanted me to stop working on the Penrose
mechanism, so that they could use my researched black hole physics, 
and results, to resurrect 
their  models of the  1970's and  early 1980's, attempting to make them work
at least as good as mine, and if that failed, just take 
mine---sooner or later.   
The literature  reveals that 
this ``plan'' went into full motion in 1999.  

What, however, probably
delayed putting this so-called plan into full motion until 1999,
other than the lack of opportunity,  is 
my continued quest to use the Penrose
produced particles and electrodynamics 
to generate the observed jets.  I decided to attend University of Florida
(UF) instead of Caltech because (1) I initially wanted to 
do my postdoctoral work at UF,  working  with the relativistic astrophysics
group there;
(2) I could continue my research on the Penrose mechanism as an 
energy source for AGNs, with the 
support of my postdoctoral advisor, Henry Kandrup: for
there was still much to be revealed about the four-momentum trajectories 
of the escaping 
Penrose particles in the Kerr metric, that needed to be investigated.
Now,  as far as my understanding
of the trajectories, I was making considerable progress, finding
that many of the features I expected to be produced by  
electrodynamics were actually being intrinsically produced by the Kerr
black hole.
But each year, from 1995, after my 
41 paper (Williams 1995) was published in PRD, 
over and over, 
I submitted follow-up
papers to ApJ and/or PRD, to share my new findings, and to improve
my credentials,  as well as  submitted proposals to NSF and NASA's grant
programs, to support my research as first a postdoctoral
associate, then an assistant scientist at UF, only
to be rejected or denied time after time.  So with no new 
refereed papers published, no funds, I was forced to accept positions 
that were not conducive
to my research, working basically in a ``vacuum'': a visiting 
assistant professorship at North Carolina A \& T (1997-1998) and an
associate professorship at Bennett College 
(1998-2002) before returning to UF, to effectively continue by research,
since both these past positions
required a lot of teaching, leaving little time for research 
projects.  
Yet, I still managed to 
present my research at some international conferences and workshops [e.g.,
Texas Symposium on Relativistic Astrophysics (1996, 2000),
 Marcel Grossman Meeting 
on General Relativity (1997), American Physical Society (1998)  
American Astronomical Society (2000), Aspen Center for Physics (2000)], 
to promote my model of AGNs, and to get some papers 
published in proceedings.  But, I had no luck
in getting my papers published in ApJ nor PRD; it seemed that both
these journals, I was told, had difficulty in finding referees
that would give a report: For example, a certain referee would 
agree, then many months later state that they cannot give a report.
The reason, I was told by editors   
of ApJ and PRD, is that the subject matter is not a 
familiar one.
Finally, when a referee would agree to give a report, it would
be extremely negative.  This happened over and over again; with
each revised improved submission, the reports got more and more 
negative.  Even when I submitted Williams (1999a) to General 
Relativity
and Gravitation (GRG) in 1997 (after the 1996 submission was rejected by PRD), 
one of the two required referees recommended publication, but unfortunately
the second referee was the same referee of my PRD submission, and his 
{\elevenit same} negative comments (even though the manuscript
had been revised), as well as his status in the astrophysics community, 
ruled; a third referee suggested that 
the paper be submitted to MNRAS or ApJ.  Eventually, the editor of GRG 
suggested I resubmit the manuscript to PRD---since this is where my
previous manuscript on the subject had been published, and so I was
right back where I started.
Yet, I was determined not to give up on my AGN model,
particularly, since new observations  were becoming more and more
consistent with my results.  I knew it would only be a matter
of time before my hard work and longsuffering would pay off. 
But then something happened in 1999.

Right around the time the panel at NSF met, in early January 1999,
 to review the fifth proposal I had submitted, another one I 
was not awarded,
although this year it  was
recommended to
``award'' me, ``this worthy Principal Investigator'' 
(see also the above comments in Section~I),
an ApJ {\elevenit Letters} paper by Krolik (1999) was
received January 13, 1999,  apparently using
black hole physics
presented in my manuscript (Williams 1995), of the orbits
inside the marginal stable orbit (or the so-called plunging
region), properties and conditions I devised, 
to work out the Penrose mechanism.
Krolik applies this physics to the BZ-type models.
His conclusion of possible magnetic torque inside the
plunging region was, with little doubt, based on my results, 
which showed that particle
orbits could exist in this region, inside the marginally stable orbit,
contrary to what was thought
previously by accretion disk theorists.  Krolik even states ``there
is the potential here for a realization of the Penrose process,'' yet,
does not
reference my work.  Moreover, in the Acknowledgements, he
thanks Mitch Begelman, Roger Blandford, and Scientific Editor
Ethan Vishniac of ApJ.  
Note, in my proposal to NSF (1999), 
I requested
funding to investigate a three-dimensional 
time-dependent evolution of the 
jets intrinsically produced by the Penrose process:  without and 
with the accretion disk magnetic field, to see if the jets 
extend to their observed distances.

In addition, similarly, after the panel met that January, an 
ApJ {\elevenit Letters}
paper by Gammie (1999), received April 6, 1999, claimed to have worked
out a model developed by Takahashi et al. 1990, the so-called
MHD Penrose process, stating that energy is being extracted in
this model, from within the ergosphere.  Not only did it appear
that Gammie used the particle orbits derived 
in Williams (1991, 1995; with the results presented again 
in Williams 1999a), 
enabling him to get 
rid of a degree of 
freedom by expressing 
$E=E(L)$, where $E$ and $L$ are the conserved energy and azimuthal
angular momentum as measured by an observer at infinity, but it seems 
he also used a modified version of my initial conditions, and 
final analytical expression
that transforms from the 
local nonrotating frame to the Boyer-Linquist frame (i.e., the 
frame of the observer at infinity), which shows the Penrose
energy extraction process.   The ``guiding center approximation''
stating that the single particle approach must be used when a
fluid, as in MHD, is in a significantly strong gravitational field, 
allowed 
him to use this final expression
(for further details see Appendix of Williams 1999b). 
Thus, it is no wonder that he
was able to extract
{\elevenit some} energy, but little---only because
of the physical way he set the problem up.  Nevertheless,
it clearly shows that
my equations work in a general sense.  Note, however, due to the 
complexity of the problem, it is literally impossible for Gammie 
to have known how to use these relations except having 
prior knowledge of them being used in my detailed calculations
(compare Williams 1995).  Now, with Gammie performing a 
one-dimensional space
Penrose analysis: within a magnetic field, would not it be appropriate
for him to reference my three-dimensional
 space Penrose analysis: without a magnetic
field? at least to compare the results.
  Yet, he does not, which is not surprising,  
for Gammie (1999)  acknowledges that he is 
grateful to both Blandford
and Krolik.
 
Concerning the above,  I am not sure who served on the NSF 1999 panel,
and may never know, or if it had anything to do with the timing, but after
January 1999 a large flux of related papers began being published 
in ApJ and PRD, altogether tied with Blandford and/or Thorne and Phinney,
with each using the particle orbits devised in Williams (1995),
applying them to the BZ-type models, more or less, 
claiming to have worked out the Penrose mechanism; or using
these orbits in gravitational wave phenomena (e.g., coalescing or
merging binaries),
claiming to have solved the three-dimensional Kerr trajectories for
$E$ and $L$, independently of my
solutions or 
numerically (see e.g., Hughes \& Blandford 2003)---anything but to 
appropriately 
reference by paper.
Strangely, out of all the recently proposed black hole models
for AGNs,  by authors
that I can  
 identify 
as using my papers, or my proposals, 
are closely affiliated with Blandford, including the work of  Meier \&
Koide et al. (2000, 2002).  This has occurred too frequently to
be a coincidence.  Even if these
authors were not aware of my work, one would expect Blandford to at least
bring it to their attention, so that they could
appropriately reference my papers.  
Similarly, Begelman, who also knows of my work 
(Williams 1995), a collaborator and friend of 
Blandford,  a co-author of the
classic paper by Wilms et al. (2001), which reports observations of 
the X-ray illumination
from within the central region of the accretion disk, suggesting that
 energy is being extracted
from a rotating black hole, mentions in this paper the BZ
model, but then associates it with the Penrose mechanism.  Yet, there is 
no mentioning of my model, which shows in details how Penrose 
Compton scattering
within the marginally stable orbit produces escaping X-rays
(Williams 1995, 1999a, 2002) that would naturally illuminate
the accretion disk, giving rise to a steep emissivity profile,
being consistent 
with these observations (Williams 2004).  

Now, the Proposed Plan of Research mentioned in Section~I, to use 
the Kerr-Newman black hole, 
 was before
finding out about the vortical escaping
Kerr orbits, in 1997,
from Fernando de Felice at the 
Marcel Grossmann meeting in Israel.  After my Parallel Session 
presentation (Williams 1999c), concerning the influence of the 
gravitomagnetic field on the trajectories of 
escaping Penrose scattered particles, he wanted to know if the particles
escape along vortical orbits. 
We talked and he directed me 
 to a paper (de Felice and Carlotto 1997) describing gravity induced 
vortical trajectories.  At the time, however, I knew that my particles 
escape 
with large 
angular and fairly large polar
coordinate momenta, with relativistic energies, but that was all,
which is why I proposed  to
do a time-dependent evolution of the escaping trajectories.
Yet, it was the 
paper by de Felice and Carlotto (1997) that led me to investigate, 
finding
that the Penrose scattered particles escape along vortical
trajectories collimated about the polar axis in the form of jets.
Such vortical trajectories are a general relativistic effect due to the 
frame dragging and the transferring
of orbital angular momentum to the scattered particles, as well as
the GM force  acting on the momentum of the particles (Williams
1999a). 
\vglue .5in
 
\centerline{\bf IV. PRESENT ERA}
\medskip      

It now appears that I have
everything needed
to generate the jets, even an intrinsically induced dynamo magnetic
field
due to the escaping plasma (i.e., charged particles), that could 
magnetically aid in confining the  particles.  
Such a magnetic field, being
similar to that of a solenoid, would be carried along with the $e^-e^+$ 
plasma jet.
This however needs further investigation. 

Importantly, in these Penrose processes we do not need the
magnetic field of the accretion disk to ``communicate'' between the
accretion disk and the
black hole.  Therefore, there is no need for the
BZ type models (and
their many associated problems) in the direct
role of energy extraction from a spinning black hole.
But the presence of BZ-type models appears to be need once particles 
are on
escaping orbits, serving the same effects they do in the jets of
protostars, i.e., seeming to have a dominant role on a large scale,
within the weak field limit,
at distances outside the strong effects of general relativity.
 
As for producing the observed
synchrotron radiation, indicating the present of a magnetic field
near the core region, such radiation
could very well be produced by the expected intrinsically self-induced
magnetic field due to the
dynamo-like
action of the escaping Penrose produced $e^-e^+$ pairs, escaping
on vortical,
coil-like  helical trajectories concentric the polar axis, in the form of a
swirling ``current'' plasma.  This, therefore, adds
more to the
unimportance  of the accretion disk magnetic field near the
event horizon.  Now, whether or not the helical jet structure of the 
escaping Penrose particles and the intrinsically induced magnetic field
are related to recent VLBA 
polarization measurements (see Homan 2004 and references therein)
remains to be seen; at least, however, the helical trajectories of the 
escaping plasma are 
consistent.
\vglue .5in

\centerline{\bf V. CONCLUSIONS} 
\medskip 

In this {\elevenit Letter}, I have described events that led to
my needing to speak out to set the record straight on 
modern black hole physics.  
In closing, I want to apologize to the innocent, i.e., the 
ones
that were unaware that it was my work being suggested to 
do by Blandford, Thorne, Phinney, Krolik, and/or Gammie. 
For example,  former Caltech graduate student Hughes, who was apparently
advised by Thorne, Blandford, and Phinney to ``rederive''
$E$ and $L$, presents his
derivations in a series of 
papers published in PRD (Hughes 2000, 2001a, 2001b), yet did not reference
my manuscript(s), even though it is quite clear that he used
analytic equations unique to my derivation presented in 
Williams (1991, 1995, 1999a).  For completion and
comparison below, 
I derived functions $E(a,r,Q)$ and $L(a,r,Q)$; Hughes derives
$L(a,r,E)$; $Q(a,r,E)$ and $E(a,r,L)$; $Q(a,r,L)$,
where $Q$ is the so-called Carter
constant, and $a$ is the angular
momentum per unit mass parameter.  
But I do not fault this former graduate student: for he was only
doing what he was advised to do.  Yet, I do fault Editor Nordstrom
of PRD for permitting it to happen, who was well aware of my work, and of
 my being Black, from my repeated manuscript submissions and rejections,
while papers of models by graduate students of Thorne, Blandford, and/or 
Phinney, using black hole physics presented in my manuscripts,
were allowed to be published.  Moreover, 
in a recent ApJ {\elevenit Letters} paper
by authors Hughes and Blandford (2003), entitled ``Black Hole
Mass and Spin Coevolution by Mergers'' in which the nonequatorial
orbital conserved energy $E(a,r,Q)$ and $L(a,r,Q)$ are needed, 
just the very functions derived in Williams (1991, 1995),
and the simplicity of taking the derivative of Equation (A13)
of Williams (1991, 1995), solving for $Q(a,r,E,L)$, yielding 
$Q(a,r)$,
one cannot refrain from questioning the truthfulness in the statement
made by these authors that the conserved quantities: $E$, $L$, $Q$
 were solved for 
numerically, particularly when we know that Blandford had 
knowledge and access of analytical solutions from my work.

Now, I know that there
were a few other scientists involved, but the  senior scientists named 
(particularly, Blandford, Thorne, Phinney, Krolik, Gammie, Vishnaic, 
Nordstrom),
based on a chain of events, played the most prominent roles. 
So, overall, 
I think I pretty much have the scenario right, i.e., as to how I believe
that my intellectual property: the knowledge (or research) of the black 
hole physics  
conveyed in my papers,  and in my various grant proposals, was 
``stolen'' from me, and given to some other scientists to use in 
their models, 
where in many cases, those scientists
have gain recognition for that very knowledge used. Thou
shalt not steal.

I realize that one
can not reference every paper one obtains knowledge from, particularly
if the material is not applied directly.  But having knowledge that I 
successfully worked out the Penrose mechanism (Williams 1991, 1995) 
and to mention the Penrose mechanism in your manuscript, 
and willfully not mention by work,
is straight-out scientifically morally wrong!  

Scientists out there that  know me, know that I am a nice
hard working person, who loves science, particularly relativistic 
astrophysics, and I do not deserve what I have been put through
 these  past years.
It has been distracting and has kept me from the science I am
capable of doing;
therefore, it must stop.  This {\elevenit Letter}, 
I am praying, shall do the job.

Finally,
we study science for the beauty of it.  It is interesting how
we are steadily, it seems, being revealed more and more of
the mysteries of the Universe and the physics of how things 
of nature work.  These revelations are being revealed, however,
to scientists that work hard to seek answers in a
righteous way.  This goes beyond the color of ones skin and the texture 
of ones hair.  Every scientist wants recognition 
for his contribution to science, just like the athlete wants to
win, yet they both enjoy ``playing the game.''  Ones ethnic race should not
exclude a scientist from being recognized for their
contributions.  So I am appealing to the scientific
community, give to me the recognition that I am due for my
contribution to the study of black hole physics.
\vglue .5in

\centerline{\bf REFERENCES}
\medskip  

\parindent=0pt

Bardeen, J. M., Press, W. H., \& Teukolsky, S. A. 1972,
ApJ, 178, 347.

Bi\v{c}\'{a}k, J. 2000, Pramana, 55, No. 4, 481
(gr-qc/0101091).

Bi\v{c}\'{a}k, J., \& Ledvinka, T. 2000, IL Nuovo Cimento,
115 B, 739 (gr-qc/0012006).

Bi\v{c}\'{a}k, J.,  \& Jani\v{s}, J. 1985, MNRAS, 212, 899.

Blandford, R. D. 1976, MNRAS, 176, 465.

Blandford, R. D., \& Payne, D. G. 1982, MNRAS, 199,
883.

Blandford, R. D., \& Znajek, R. L. 1977, MNRAS, 179, 433.

Burns, M. L., \& Lovelace, R. V. E. 1982, ApJ,
262, 87.
 
de Felice, F., \& Carlotto, L. 1997, ApJ, 481, 116.

Eardley, D. M., \& Lightman, A. P. 1975, ApJ,
200, 187.      

Eilek, J. A. 1980, ApJ, 236, 664.

Eilek, J. A., \& Kafatos, M. 1983, ApJ,
271, 804.

Gammie, C. F. 1999, ApJ, 522, L57.

Homan, D. C. 2003, in Future Directions in High Resolution Astronomy: The
10th Anniversary of the VLBA, ASP Conference Series, eds. 
J. D. Romney \& M. J. Reid, in press (astro-ph/0401320). 

Hughes, S. A., \& Blandford, R. D. 2003, ApJ, 585, L101. 

Junor, W., Biretta, J. A., \& Livio, M. 1999, Nature, 401, 891. 

Kafatos, M. 1980, ApJ, 236, 99.
 
Kafatos, M., \& Leiter, D. 1979, ApJ, 229, 46.

Koide, S., Kazunari, S., Takahiro, K., \& Meier, D. 2002,
Science, 295, 1688.

Koide, S., Meier, D. L., Shibata, K., \& Kudoh, T.
 2000, ApJ, 536, 668.

Krolik, J. H. 1999, ApJ, 515, L73.                
                                                           
Leiter, D., \& Kafatos, M. 1978, ApJ, 226, 32. 

Lovelace, R. V. E. 1976, Nature, 262, 649.

Novikov, I. D., \& Thorne, K. S. 1973, in Black Holes,
ed. C. DeWitt \& B. S. DeWitt (New York: Gordon and Breach
Science Publishers), 343.

Penrose, R. 1969, Rivista Del Nuovo Cimento: Numero
Speciale, 1, 252.

Piran, T., \& Shaham, J. 1977a, Phys. Rev. D,
16, No. 6, 1615.
 
--------. 1977b, ApJ, 214, 268.
 
Piran, T., Shaham, J., \& Katz, J. 1975, ApJ, 196, 107.   

Punsly, B. 1991, ApJ, 372, 424.

Takahashi, M., Nitta, S., Tatematsu, Y., \& Tomimatsu, A.
1990, ApJ, 363, 206.

Weinberg, S. 1972, Gravitation and Cosmology: Principles and Applications
of the General Theory of Relativity (New York: John Wiley \& Sons).

Williams, R. K. 1991, Ph.D. thesis, Indiana Univ.
 
--------. 1995, Phys. Rev. D, 51, No. 10, 5387.

--------. 1999a, 1997, 1996, Phys. Rev. D, submitted (astro-ph/0203421).
 
--------. 1999b, 1995,  ApJ, submitted (astro-ph/0306135).  
                             
--------. 1999c, in The Proceedings of The Eighth
Marcel Grossmann
Meeting on General Relativity, ed. T. Piran \& R. Ruffini
(Singapore: World Science), 416.    
 
--------. 2001, in  Relativistic Astrophysics: 20th
Texas Symposium,
ed. J. C. Wheeler \& H. Martel
(New York: American Institute of Physics),
448 (astro-ph/0111161).

--------. 2002,  ApJ, submitted (astro-ph/0210139).  

--------. 2004, in preparation.

Wilms, J., Reynolds, C. S.,
Begelman, M. C., Reeves, J.,
Molendi, S., Staubert, R., \& Kendziorra, E. 2001,
MNRAS, 328, L27 (astro-ph/0110520).

\bye